# Real-time Predictive Analytics for Improving Public Transportation Systems' Resilience

Peyman Noursalehi
Northeastern University
Boston, MA, USA
Noursalehi.p@husky.neu.edu

Haris N. koutsopoulos
Northeastern University
Boston, MA, USA
h.koutsopoulos@neu.edu

## ABSTRACT

Public transit systems are a critical component of major metropolitan areas. However, in the face of increasing demand, most of these systems are operating close to capacity. Under normal operating conditions, station crowding and boarding denial are becoming a major concern for transit agencies. As such, any disruption in service will have even more severe consequences, affecting huge number of passengers. Considering the aging infrastructure of many large cities, such as New York and London, these disruptions are to be expected, amplifying the need for better demand management and strategies to deal with congested transit facilities. Opportunistic sensors such as smart cards (AFC), automatic vehicle location systems (AVL), GPS, etc. provide a wealth of information about system's performance and passengers' trip making patterns. We develop a hybrid data/model-driven decision support system, using real-time predictive models, to help transit operators manage and respond proactively to disruptions and mitigate consequences in a timely fashion. These models include station arrival and origin-destination predictions in real-time to help transit agencies, and predictive information systems for assisting passengers' trip making decisions.

## 1. INTRODUCTION

Population of major metropolitan areas has witnessed a steady increase over the past few decades, and the trend is expected to continue. Therefore, more and more people rely on public transit systems for their daily travels. Figure 1 (*1*) shows forecasts of passenger journeys in London for different lines, with the London Overground almost doubling in ridership over the next 10 years.

However, prohibitive costs of adding new lines or increasing train frequency have hindered transit agencies efforts to maintain and improve performance levels. In a recent survey of London rail passengers (*1*), reduction in delays and cancelations (43%), increasing the capacity of trains (30%), and more frequent trains (30%) were on the top of passengers demands. This shows the major challenges faced by transit agencies. Development of effective demand management and operations control strategies, based on predictive models that use real-time data, is a promising approach. In this paper, we develop a methodology and show its potential applications in improving transit systems' performance

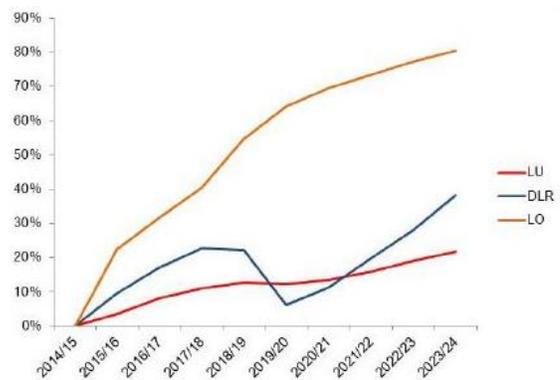

Figure 1. Forecast growth in passenger journeys

## 2. DEMAND ANALYTICS

Automatic Fare Collection (AFC) systems provide massive, detailed information on passengers' trip making and travel behaviour. In this research we develop demand management decision support systems and advanced passenger information services, utilizing these data. Three categories of predictive models comprise these systems:

- a. Predicted arrivals at each station
- b. Origin-destination predictions
- c. Predicted load for each train

The prediction time horizon varies depending on the application. Here, we focus on predictions 15 to 30 minutes ahead, which gives realistic opportunity to decide and implement a strategy in real-time. Figures 2.a and 2.b demonstrate the overall schema of offline training and online prediction steps. In the training stage, similar daily arrivals at each station are clustered in a group. Then, a separate predictive model is estimated for each group. Information exogenous to the system, such as whether or not a major event was happening at a nearby location, weather conditions and planned station or line closures, are separately put into their own categories. They enter as explanatory variables in the predictive models for each station, hence it would be possible to predict the effect of event categories on demand. In real-time as the AFC data is coming in, a classifier decides which one the patterns in the historical database is closest to the incoming data, and selects the corresponding model for the prediction task. This model would take as input any information about events that are known to happen on that day (e.g. soccer game, etc.) and includes their effect on prediction. In addition, any unplanned events, such as station closures or major road closures, can also enter the model as explanatory variables as soon as they are noticed.

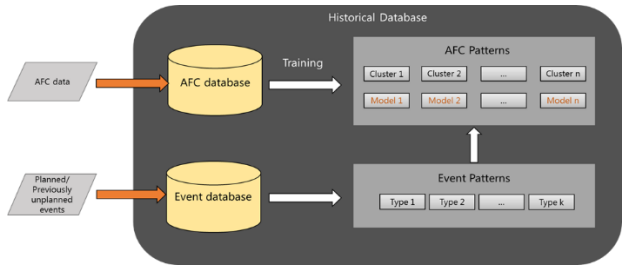

Figure 2.a Offline Training

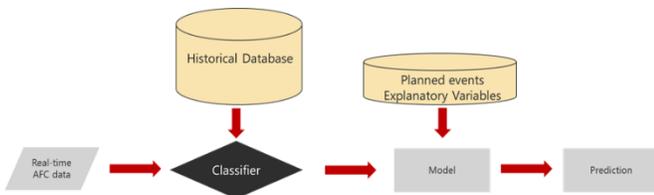

Figure 2.b Real-time Predictions

Additionally, many transit agencies have (and often make available through APIs) train positions. Coupled with the arrivals and origin-destination predictive models, it is possible to predict expected trains' loads. This information is subsequently used to forecast crowding hotspots, and evaluate corrective actions based on predicted impacts.

## 3. DECISION SUPPORT TOOLS IMPLEMENTATION

Figure 3 shows the implementation of such a decision support system. As an example, at 8 am the system takes as input, data on AFC transactions up to that point, along with information about whether any major events are happening or will happen in the next 15 or 30 minutes. Then it provides forecasts on number of arrivals at each station and their destination distribution, at 8:15 (1-step) and 8:30 (2-steps) am. Predictions are adaptive. After 15 minutes (or chosen time step), when new observations become available. The models will re-adjust their estimate of system's state based on the difference between their predictions and observed demand. New predictions are then made for 8:30 and 8:45 am, and the process continues. As a case study, Oyster transactions at the London Underground (LU) were used. This data provides both station entries and exits for February of 2013.

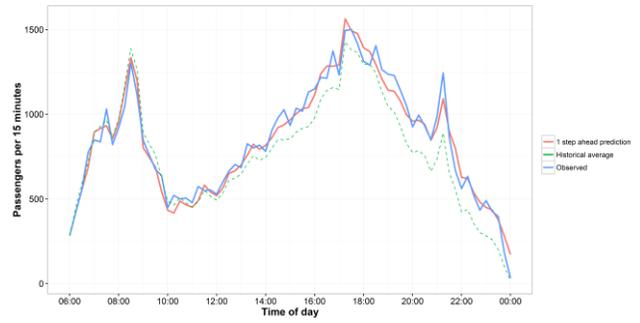

Figure 4.a 1-step ahead predictions

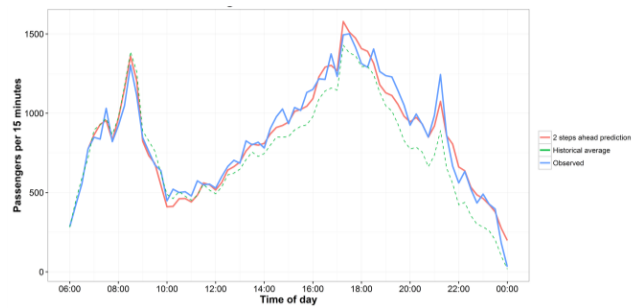

Figure 4.b 2-steps ahead predictions

State-space time series models were estimated for modeling arrivals at each station. 1-step and 2-steps ahead predictions, along with observed and historical average demand for a station are shown in Figure 4.a and 4.b. It can be seen that predictions (red)

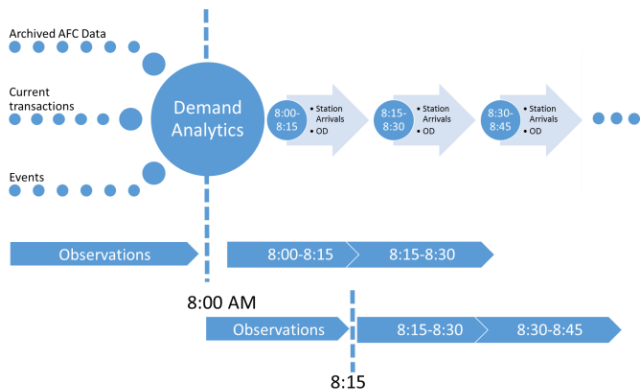

Figure 3

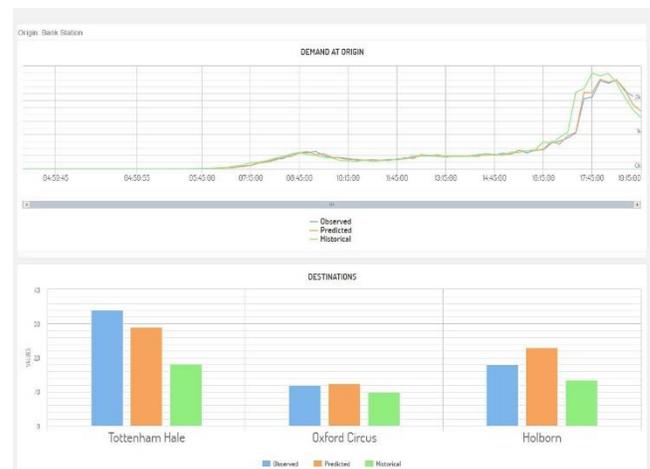

Figure 5 Real-time visualization of origin-destination demand

quickly adapt to deviations of observed demand (blue) from historical averages (green).

Interactive visualization dashboards have been developed to complement the analytical methods and communicate the outputs. Figure 5 shows a real-time dashboard for a sample of stations. The upper panel shows arrivals predictions, along with observed and historical averages for the origin station. The bottom panels show destination predictions of these passengers (only a subset of destinations are shown).

Demand prediction models serve as engine of the predictive decision support system (Figure 6). Using historical and real-time

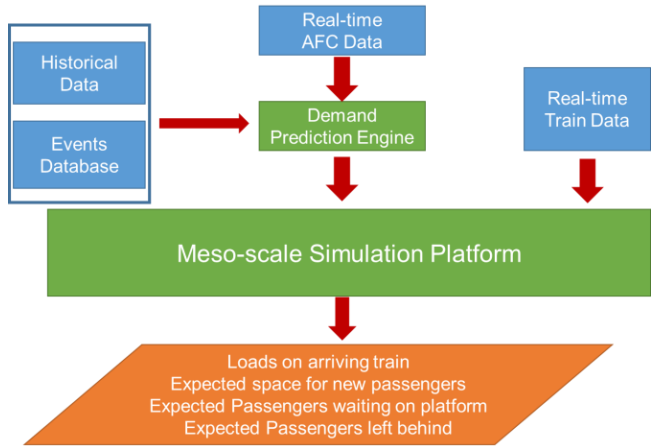

Figure 6. Platform for predictive decision support

AFC transactions, every 15 minutes they output origin-destination predictions for the next 15 or 30 minute period, for all stations. Then, a meso-scale simulation platform, simulates passenger arrivals, as well as train positions for the same period. The output of simulation includes expected loads on each train, expected number of passengers waiting on platform, and expected number of passengers who won't be able to board a train. This simulation is always running in the background, and uses real-time train data to adjust its predictions of train loads and station crowding. A prototype of this predictive decision support platform has been developed for Central Line in London, and will be extended to include all major lines of London Underground.

## 4. APPLICATIONS
### 4.1. PROACTIVE CROWD MANAGEMENT
Overcrowding at stations is a major problem with busy subway systems with significant impact on operations, passengers, and safety. During rush hours, an often practiced crowd management strategy is to close station gates at busy stations, denying passenger entrance. This results in long waiting times for large number of passengers. Some of them move to nearby stations, which in turn may experience deteriorating performance as well. A more effective and proactive strategy would be to identify opportunities to briefly close gates at stations upstream the line to help manage crowding at busier stations that are predicted to experience crowding in the next 15 or 30 minutes. Additionally, it would provide the opportunity to alert the staff earlier to implement safety measures to manage increased customer demand.

### 4.2. PASSENGER INFORMATION SERVICES
Passengers may change their trip making decisions upon timely information about upcoming gate closures or service disruptions, in real-time. Information from predictive models can be communicated to users in a number of ways:

a. Provide information about projected entries at stations and probability of entrance denial;
b. Provide personalized messages to warn registered passengers with targeted information that their journey may be disrupted and suggest alternative routes specific to them (*2*);
c. Reach with specific information only customers expecting to be travelling to / from affected stations while the disruption is ongoing.

## 5. CONCLUSION
We presented a hybrid data/model-driven decision support system, which relies on predictive models to help practitioners manage and respond proactively to demand volatility and service disruptions. A set of visualization dashboards were developed to communicate model outputs effectively and easily. This platform can serve as a crucial building block form applications that help users with trip planning or decision making under disruptions.